\documentclass[aps,showpacs,preprintnumbers,amsmath,amssymb]{revtex4}
\usepackage{dcolumn}
\usepackage[dvips]{epsfig}
\usepackage{graphicx}
\usepackage{amssymb}
\usepackage{color}
\usepackage{enumerate}
\usepackage{subfigure}
\newcommand{\degree}{^\circ}

\begin{document}
\baselineskip=0.8 cm
\title{ Shadow casted by a Konoplya-Zhidenko rotating non-Kerr black hole}
\author{Mingzhi Wang$^{1}$,  Songbai Chen$^{1,2,3}$\footnote{Corresponding author: csb3752@hunnu.edu.cn}, Jiliang
Jing$^{1,2,3}$ \footnote{jljing@hunnu.edu.cn}}
\affiliation{$ ^1$Institute of Physics and Department of Physics, Hunan
Normal University,  Changsha, Hunan 410081, People's Republic of
China \\ $ ^2$Key Laboratory of Low Dimensional Quantum Structures \\
and Quantum Control of Ministry of Education, Hunan Normal
University, Changsha, Hunan 410081, People's Republic of China\\
$ ^3$Synergetic Innovation Center for Quantum Effects and Applications,
Hunan Normal University, Changsha, Hunan 410081, People's Republic
of China}

\begin{abstract}
\baselineskip=0.6 cm
\begin{center}
{\bf Abstract}
\end{center}

 We have investigated the shadow of a Konoplya-Zhidenko rotating non-Kerr black hole with an extra deformation parameter. The spacetime structure arising from the deformed parameter affects sharply the black hole shadow.
 With the increase of the deformation parameter, the size of the shadow of  black hole increase and its shape becomes more rounded for arbitrary rotation parameter. The D-shape shadow of black hole emerges only in the case $a<\frac{2\sqrt{3}}{3}M$ with the proper deformation parameter.
 Especially, the black hole shadow possesses a cusp shape with small eye lashes in the cases with $a>M$, and the shadow becomes less cuspidal with the increase of the deformation parameter. Our result show that the presence of the deformation parameter yields a series of significant patterns for the shadow casted by a Konoplya-Zhidenko rotating non-Kerr black hole.

\end{abstract}
\maketitle

\newpage
\section{Introduction}

The increasing astronomical observation support the existence of  supermassive black holes at the center of many galaxies. Therefore, measuring black hole parameters becomes very significant since it can help us to understand features of black hole and to examine further various theories of gravity. Black hole shadow is a two-dimensional dark zone in the observer's sky where light from a source is captured by the black hole. It is shown that the shadow can be treated as a useful tool of detecting  black hole parameters because its shape and size  carry the fingerprint of the geometry around the black hole \cite{sha1,sha2,sha3}. For example, the shadow for a static black hole is a perfect circle, while for the rotating Kerr one, it becomes an elongated silhouette in the direction of the rotation axis due to the dragging effect \cite{sha2,sha3}. Moreover, the effects of other characterizing  parameters on black hole shadow have been studied in the last few years \cite{sha4,sha5,sha6,sha7,sha9,sha10,sha11,sha12,sha13,sha14,sha15,sha16,
sb1,sha17,sha18,sha19}, which indicate that these  parameters imprint the shape and size of the shadow. Especially, in the cases where the null geodesics are not variable-separable, it is found that the black hole shadow possesses some novel features. For a Kerr black hole with scalar hair \cite{sw,swo,astro,chaotic} or a binary black hole system \cite{binary}, there exist fractal structures in the black hole shadow due to chaotic motion of photons. Moreover, for a Kerr black hole with Proca hair \cite{harip1}, the black hole shadow has a cusp silhouette as the black hole parameters lie in a certain range \cite{fpos2}. The further analysis show that these novel patterns in shadows are related to the non-planar bound photon orbits, which are also called as the fundamental photon orbits (FPOs) \cite{fpos2}. In order to disclose entirely the characteristics of the shadows, it is necessary to study further the shadow of black holes in various theories of gravity.

Einstein's General relativity is considered probably the most beautiful of all existing physical theories, which has successfully passed a series of observational and experimental tests \cite{t3}. However, the current observations cannot completely exclude the possibility of the deviation from Einstein's gravity theory, which leaves an ample room for other alternative theories of gravity. Recently, Konoplya and Zhidenko \cite{kz} have proposed recently a rotating non-Kerr black hole metric beyond General Relativity through adding a static deformation,  which can be looked as an axisymmetric vacuum solution of a unknown alternative theory of gravity \cite{RLs}. The Konoplya-Zhidenko rotating non-Kerr black hole has three parameter, i.e.,  the  mass $M$, the rotation parameter $a$, and the deformation parameter $\eta$.  The extra deformation parameter $\eta$ describes the deviation
from the usual Kerr one and modifies sharply the structures of spacetime in the strong-field region\cite{kz,sy}. Making use of this rotating non-Kerr metric, they found that there exist some non-negligible deviation from the Kerr spacetime which lead to the same frequencies of the black-hole ringing \cite{kz}. Moreover, the examinations from the iron line \cite{GKt02} and the quasi-periodic oscillations \cite{GKt01} also endorse that the geometry of
a real astrophysical black hole could be described by such a rotating non-Kerr metric.

Since the shadow of black hole is determined by the propagation of light ray in the spacetime, it is expectable that the new properties of spacetime structure originating from the deformation parameter will yields some new effects on the black hole shadow. Therefore, in this paper, we focus on the investigation of the shadow casted by a Konoplya-Zhidenko rotating non-Kerr black hole and then probe the signature of the deformation parameter resides in the black hole shadow.

The paper is organized as follows. In Sec. II, we review briefly the metric of the Konoplya-Zhidenko rotating non-Kerr black hole  and then analyze the propagation of light ray in this background. In Sec. III, we investigate the shadows casted by Konoplya-Zhidenko rotating non-Kerr black hole. Finally, we present a summary.

\section{The photon orbit in the Konoplya-Zhidenko rotating non-Kerr black hole spacetime}

Firstly, let us review briefly the Konoplya-Zhidenko rotating non-Kerr metric obtained in Ref.\cite{kz}. As a usual rotating non-Kerr case, it describes the geometry of a rotating black hole with the deviations from the Kerr one through adding an extra deformation. In the Boyer-Lindquist coordinates, the metric has a form \cite{kz}
\begin{eqnarray}
\label{xy}
ds^{2} &=& -\bigg(1-\frac{2Mr^2+\eta}{r\rho^{2}}\bigg)dt^{2}
+\frac{\rho^{2}}{\Delta}dr^{2}+\rho^{2} d\theta^{2}+\sin^{2}\theta\bigg[r^{2}+a^{2}
+\frac{(2Mr^{2}+\eta)a^2\sin^{2}\theta}{r\rho^{2}}\bigg]d\phi^{2}\\ \nonumber
&-&\frac{2(2Mr^2+\eta)a\sin^{2}\theta}{r\rho^{2}}dtd\phi,
\end{eqnarray}
with
\begin{equation}
\Delta=a^{2}+r^{2}-2Mr-\frac{\eta}{r},\;\;\;\;\;\;\;\;\;\; \rho^{2}=r^{2}+a^{2}\cos^{2}\theta.
\end{equation}
where $M$, $a$ and $\eta$ denote the mass, the angular momentum and the deformation parameter of black hole, respectively. The deformation parameter $\eta$  describes the deviations from the Kerr metric. As the parameter $\eta$ vanishes, the metric reduces to that of usual Kerr spacetime. Comparing with the Kerr black hole,  the presence of the deformation parameter extends the allowed range of the rotation parameter $a$ and  changes the spacetime structure of the black hole in the strong field region \cite{kz,sy}.
In the case $a<M$, the condition for the existence of black hole horizon in this spacetime becomes \cite{sy}
\begin{equation}
\label{hcz}
\eta\geq \eta_{c1}\equiv-\frac{2}{27}(\sqrt{4M^{2}-3a^{2}}+2M)^{2}(\sqrt{4M^{2}-3a^{2}}-M),
\end{equation}
while in the case $a>M$, it becomes $\eta>0$. When $\eta$ and $a$ lie in other regions, there is no horizon and then the spacetime (\ref{xy}) becomes a naked singularity. Considering the weak cosmic censorship conjecture,  we here will study only the case where there exists horizon and the metric (\ref{xy}) describes the gravity of a  black hole.
The value of $\eta$ determines the number and positions of black hole horizons.  These spacetime properties affect the propagation of photon and further changes  shadow of a Konoplya-Zhidenko rotating non-Kerr black hole.

The Hamiltonian of a photon propagation along null geodesics in a Konoplya-Zhidenko rotating non-Kerr black hole spacetime can be expressed as \cite{hamin}
\begin{equation}
\label{hami}
H(x,p)=\frac{1}{2}g^{\mu\nu}(x)p_{\mu}p_{\nu}=0.
\end{equation}
Since there exist two ignorable coordinates $t$ and $\phi$ in the above Hamiltonian, it is easy to obtain two conserved quantities $E$ and $L_{z}$ with the following forms
\begin{eqnarray}
\label{EL}
E=-p_{t}=-g_{tt}\dot{t}-g_{t\phi}\dot{\phi},\;\;\;\;\;\;\;\;\;\;\;\;\;\;
L_{z}=p_{\phi}=g_{\phi\phi}\dot{\phi}+g_{\phi t}\dot{t},
\end{eqnarray}
which correspond to the energy  and the $z$-component of the angular momentum
of photon moving in the Konoplya-Zhidenko rotating non-Kerr black hole spacetime. With these two conserved quantities, the null geodesic equation can be written as
\begin{eqnarray}
\label{tfc}
\dot{t}&=&E+\frac{(a^{2}E-aL_{z}+Er^{2})(2Mr^{2}+\eta)}{\Delta\rho^{2}r},
\\
\label{jfc}
\dot{\phi}&=&\frac{aE\sin^{2}\theta(2Mr^{2}+\eta)
+a^{2}L_{z}r\cos^{2}\theta-L_{z}(2Mr^{2}
-r^{3}+\eta)}{\Delta\rho^{2}r\sin^{2}\theta},\\
\label{rfc}
\rho^{4}\dot{r}^{2}&=&R(r)=-\Delta[Q+(aE-L_{z})^{2}]+[aL_{z}-(r^{2}+a^{2})E]^{2},
\\
\label{thfc}
\rho^{4}\dot{\theta}^{2}&=&p_{\theta}^{2}=Q-\cos^{2}\theta
\bigg(\frac{L_{z}^{2}}{\sin^{2}\theta}-a^{2}E^{2}\bigg),
\end{eqnarray}
where the quantity $Q$ is the generalized Carter constant related to the constant of separation $K$ by $Q=K-(aE-L_{z})^2$. The constant $K$ is associated with the hidden symmetries of the spacetime generated by a second order Killing tensor $K^{\mu\nu}$ . The Killing tensor satisfies the equation
\begin{eqnarray}\label{killing}
\nabla_{(\lambda}K_{\mu\nu)}=0,
\end{eqnarray}
where $\nabla$ denotes the covariant differentiation with respect to the spacetime metric $g^{\mu\nu}$ and the round brackets denote symmetrization over the indices enclosed. With the Killing tensor $K^{\mu\nu}$, the constant of separation $K$  can be expressed as $K=K^{\mu\nu}p_{\mu}p_{\nu}$.
In Boyer-Lindquist coordinates, one can find that the Killing tensor satisfied equation (\ref{killing}) in the Konoplya-Zhidenko rotating non-Kerr black hole spacetime has a form
\begin{eqnarray}
K^{\mu\nu}=\frac{(r^2+a^2)^2}{\Delta}\delta^{\mu}_t\delta^{\nu}_t
-\Delta\delta^{\mu}_r\delta^{\nu}_r+\frac{(r^2+a^2)a}{\Delta}\delta^{\mu}_t\delta^{\nu}_{\phi}+
\frac{a^2}{\Delta}\delta^{\mu}_{\phi}\delta^{\nu}_{\phi}+r^2g^{\mu\nu}.
\end{eqnarray}
With help of the null tetrad,
\begin{eqnarray}
l^{\mu}=\bigg(\frac{r^2+a^2}{\Delta},1,0,\frac{a}{\Delta}\bigg),\;\;\;\;\;\;\;\;\;\;\;\;\;\;
n^{\mu}=\bigg(\frac{r^2+a^2}{2\rho^2},-\frac{\Delta}{2\rho^2},0,\frac{a}{2\rho^2}\bigg),
\end{eqnarray}
the components of the Killing tensor $K^{\mu\nu}$  can be written as
\begin{eqnarray}
K^{\mu\nu}=2\rho^2l^{(\mu}n^{\nu)}+r^2 g^{\mu\nu},
\end{eqnarray}
which is similar to that for the ordinary Kerr spacetime \cite{hamin1}. 

The unstable spherical orbits are very important to determine the boundary of the shadow by a black hole. The spherical orbits satisfy
\begin{eqnarray}
\dot{r}=0,\;\;\;\;\; \text{and} \;\;\;\; \ddot{r}=0,
\end{eqnarray}
which yield
\begin{eqnarray}
\label{r}
R(r)&=&-\Delta[Q+(aE-L_{z})^{2}]+[aL_{z}-(r^{2}+a^{2})E]^{2}=0,
\\
\label{r1}
R'(r)&=&-4Er[aL_{z}-(r^{2}+a^{2})E]-[Q+(aE-L_{z})^{2}]
(-2M+2r+\frac{\eta}{r^{2}})=0.
\end{eqnarray}
For the unstable spherical orbits, we have
\begin{eqnarray}
\label{r2}
R''(r)=8E^{2}r^{2}-4E[aL_{z}-(r^{2}+a^{2})E]-2[Q+(aE-L_{z})^{2}]
(1-\frac{\eta}{r^{3}})>0.
\end{eqnarray}
Solving the two equation (\ref{r}) and (\ref{r1}), we find that for the spherical orbits motion of photon the reduced constants $\xi$ and $\sigma$
have the form
\begin{eqnarray}
\label{pj}
\xi&\equiv&\frac{L_{z}}{E}=\frac{2a^{2}Mr^{2}-a^{2}\eta+2\Delta r^{3}-2Mr^{4}-3\eta r^{2}}{a(2Mr^{2}-2r^{3}-\eta)},\\
\label{q}
\sigma&\equiv&\frac{Q}{E^{2}}=\frac{-r^{4}[(6Mr^{2}-2r^{3}+5\eta)^{2}-8a^{2}(2Mr^{3}+3\eta r)]}{a^{2}(2r^{3}-2Mr^{2}+\eta)^{2}}.
\end{eqnarray}
From Eq. (\ref{thfc}), we find that $\xi$ and $\sigma$ obey
\begin{eqnarray}
\label{fw}
\sigma-\xi^{2}\cot^2\theta+a^2\cos^{2}\theta\geq0.
\end{eqnarray}

\section{Shadows of a Konoplya-Zhidenko rotating non-Kerr black hole}

\begin{figure}
\center{\includegraphics[width=8cm ]{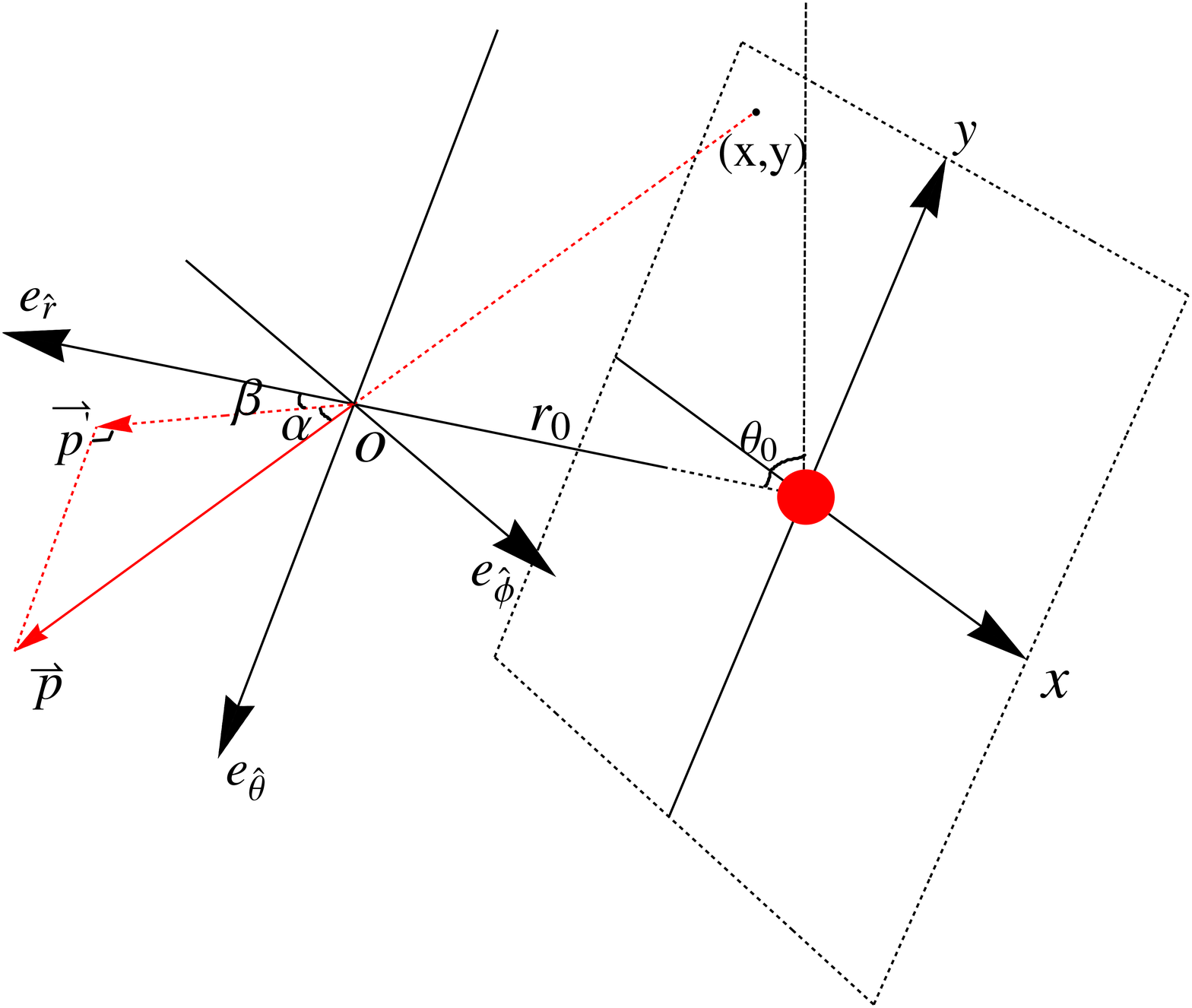}
\caption{Perspective drawing of the geometric projection of the photon's linear momentum $\vec{p}$ in
the observer's frame $\{e_{\hat{r}},e_{\hat{\theta}},e_{\hat{\phi}}\}$. The red sphere and the point $ O (r_{0},\theta_{0},0)$ denote
the position of  black hole and observer, respectively. The vector $\vec{p'}$ is the projection of $\vec{p}$ onto plane $(e_{\hat{r}},e_{\hat{\phi}})$ and  $\alpha$ is the angle between $\vec{p'}$ and plane $(e_{\hat{r}},e_{\hat{\phi}})$, $\beta$ is the angle between $\vec{p'}$ and basis $e_{\hat{r}}$.}
\label{zb}}
\end{figure}

Since the Konoplya-Zhidenko rotating non-Kerr black hole spacetime is asymptotic flat, we can define the same observer's sky at spatial infinite as in the Kerr case. The observer basis $\{e_{\hat{t}},e_{\hat{r}},e_{\hat{\theta}},e_{\hat{\phi}}\}$ can be expanded in the coordinate basis $\{ \partial_t,\partial_r,\partial_{ \theta},\partial_{\phi} \}$  as a form \cite{sha2,swo,sha18,zero1,swo7}
\begin{eqnarray}
\label{zbbh}
e_{\hat{\mu}}=e^{\nu}_{\hat{\mu}} \partial_{\nu},
\end{eqnarray}
where $e^{\nu}_{\hat{\mu}}$ satisfies $g_{\mu\nu}e^{\mu}_{\hat{\alpha}}e^{\nu}_{\hat{\beta}}
=\eta_{\hat{\alpha}\hat{\beta}}$, and $\eta_{\hat{\alpha}\hat{\beta}}$ is the usual Minkowski metric. The decomposition (\ref{zbbh}) satisfied the spatial rotations and Lorentz boosts, is not unique. Generally, it is convenient to choice a decomposition connected to a reference
frame with zero axial angular momentum in relation to spatial infinity, which is given by \cite{sha2,swo,sha18,zero1,swo7}
\begin{eqnarray}
\label{zbbh1}
e^{\nu}_{\hat{\mu}}=\left(\begin{array}{cccc}
\zeta&0&0&\gamma\\
0&A^r&0&0\\
0&0&A^{\theta}&0\\
0&0&0&A^{\phi}
\end{array}\right).
\end{eqnarray}
Here $\zeta$, $\gamma$, $A^r$, $A^{\theta}$, and $A^{\phi}$ are real coefficients.
Since the observer basis has a Minkowski normalization
\begin{eqnarray}
e_{\hat{\mu}}e^{\hat{\nu}}=\delta_{\hat{\mu}}^{\hat{\nu}},
\end{eqnarray}
one can obtain
\begin{eqnarray}
\label{xs}
&&A^r=\frac{1}{g_{rr}},\;\;\;\;\;\;\;\;\;\;\;\;\;\;\;\;
A^{\theta}=\frac{1}{g_{\theta\theta}},\;\;\;\;\;\;\;\;\;\;\;\;\;\;\;
A^{\phi}=\frac{1}{g_{\phi\phi}},\nonumber\\
&&\zeta=\sqrt{\frac{g_{\phi \phi}}{g_{t\phi}^{2}-g_{tt}g_{\phi \phi}}},\;\;\;\;\;\;\;\;\;\;\;\;\;\;\;\;\;\;\;\; \gamma=-\frac{g_{t\phi}}{g_{\phi\phi}}\sqrt{\frac{g_{\phi \phi}}{g_{t\phi}^{2}-g_{tt}g_{\phi \phi}}}.
\end{eqnarray}
Thus, the locally measured four-momentum $p^{\hat{\mu}}$ of a photon can be obtained by the projection of its four-momentum $p^{\mu}$  onto $e_{\hat{\mu}}$,
\begin{eqnarray}
\label{dl}
p^{\hat{t}}=-p_{\hat{t}}=-e^{\nu}_{\hat{t}} p_{\nu},\;\;\;\;\;\;\;\;\;
\;\;\;\;\;\;\;\;\;\;\;p^{\hat{i}}=p_{\hat{i}}=e^{\nu}_{\hat{i}} p_{\nu},
\end{eqnarray}
which means that the locally measured four-momentum $p^{\hat{\mu}}$ can be further written as \cite{sha2,swo,sha18,zero1,swo7}
\begin{eqnarray}
\label{kmbh}
p^{\hat{t}}&=&\zeta E-\gamma p_{\phi},\;\;\;\;\;\;\;\;\;\;\;\;\;\;\;\;\;\;\;\;p^{\hat{r}}=\frac{1}{\sqrt{g_{rr}}}p_{r} ,\nonumber\\
p^{\hat{\theta}}&=&\frac{1}{\sqrt{g_{\theta\theta}}}p_{\theta},
\;\;\;\;\;\;\;\;\;\;\;\;\;\;\;\;\;\;\;\;\;\;
p^{\hat{\phi}}=\frac{1}{\sqrt{g_{\phi\phi}}}p_{\phi}.
\end{eqnarray}
The spatial position of observer in the black hole spacetime is set to ($r_{o},\theta_{o},0$) as shown  in Fig. (\ref{zb}).  The $3-$vector $\vec{p}$  is the photon's linear momentum with components $p_{\hat{r}}$, $p_{\hat{\theta}}$ and $p_{\hat{\phi}}$
in the orthonormal basis $\{e_{\hat{r}},e_{\hat{\theta}},e_{\hat{\phi}}\}$,
\begin{eqnarray}
\vec{p}=p^{\hat{r}}e_{\hat{r}}+p^{\hat{\theta}}
e_{\hat{\theta}}+p^{\hat{\phi}}e_{\hat{\phi}}.
\end{eqnarray}
According to the geometry of the photon's detection, we have
\begin{eqnarray}
\label{fl}
p^{\hat{r}}&=&|\vec{p}|\cos\alpha\cos\beta, \nonumber\\
p^{\hat{\theta}}&=&|\vec{p}|\sin\alpha, \nonumber\\
p^{\hat{\phi}}&=&|\vec{p}|\cos\alpha\sin\beta.
\end{eqnarray}
Actually, the angular coordinates $(\alpha, \beta)$ of a point in the observer's local sky define the direction of the associated light ray and
establish its initial conditions. The coordinates $(x,y)$ of a point in the observer's local sky are related to its  angular coordinates $(\alpha, \beta)$  by \cite{sha2,swo,sha18,zero1,swo7}
\begin{eqnarray}
\label{xd}
x&=&-r_{0}\tan\beta=-r_{0}\frac{p^{\hat{\phi}}}{p^{\hat{r}}}, \nonumber\\
y&=&r_{0}\frac{\tan\alpha}{\cos\beta}=r_{0}\frac{p^{\hat{\theta}}}{p^{\hat{r}}}.
\end{eqnarray}
The image of a black hole shadow in observer's sky is composed of the pixels corresponding the light rays falling down into the black hole horizon. The unstable spherical orbits of photons provide us the boundary of the shadow.

\begin{figure}
\includegraphics[width=7cm ]{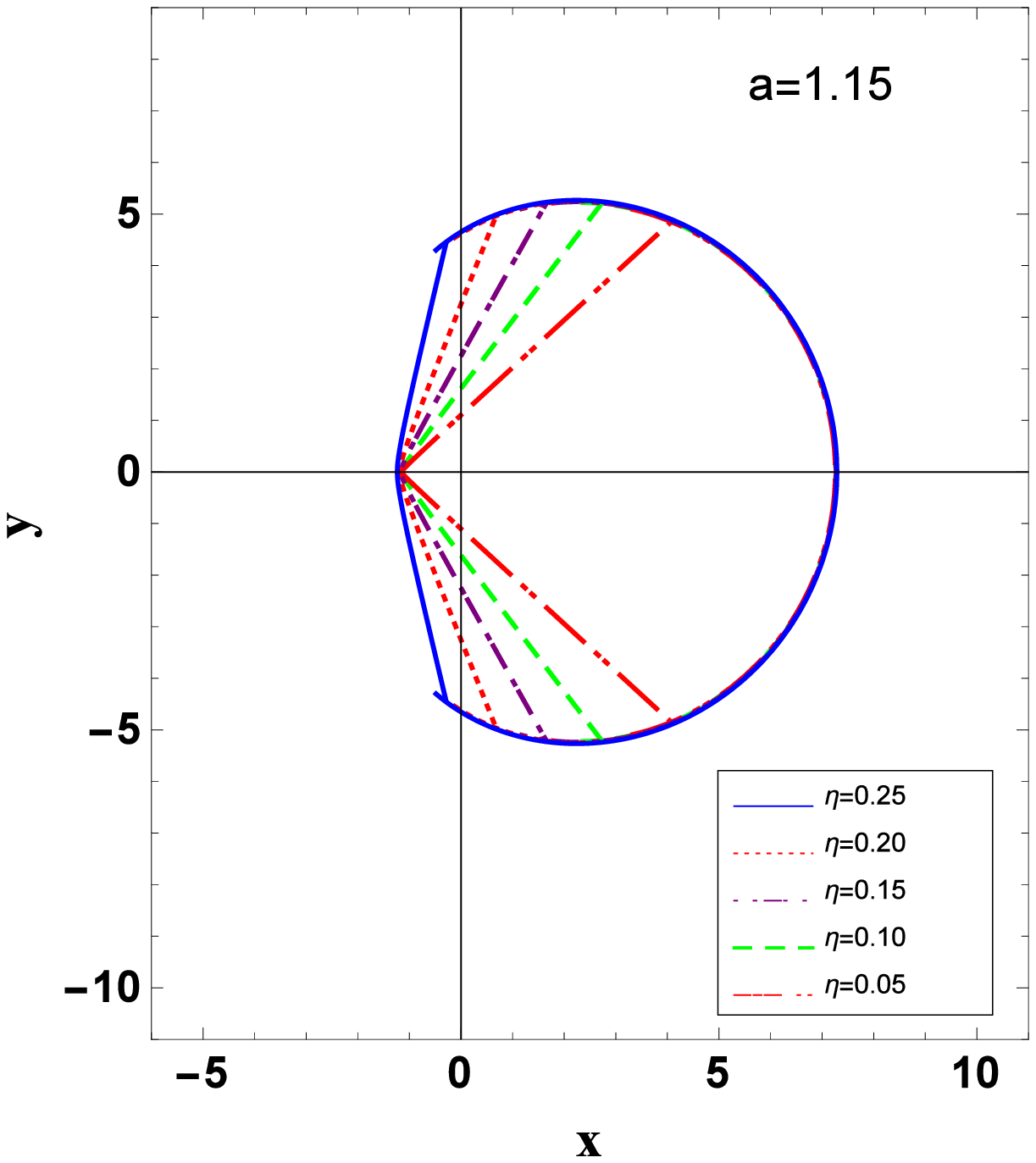}\includegraphics[width=7cm ]{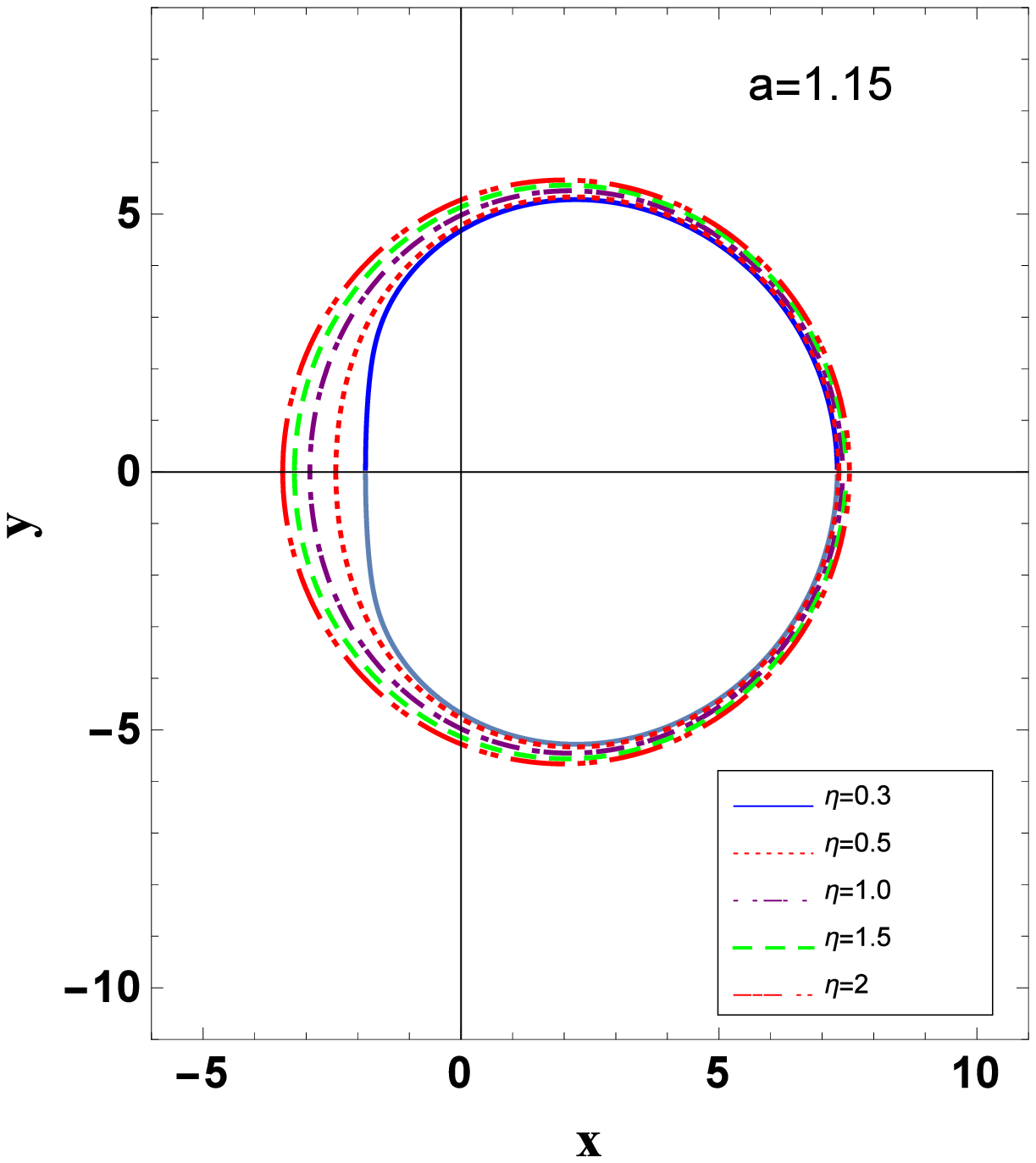}
\caption{The shadow of a Konoplya-Zhidenko rotating non-Kerr black hole with deformation parameter $\eta$ for the fixed $a=1.15$ lied in the range $M<a<\frac{2\sqrt{3}}{3}M$.  Here, an observer is situated at the origin of coordinates with the inclination angle $\theta_{0}=90\degree$.}
\label{a050}
\end{figure}

\begin{figure}
\includegraphics[width=7cm ]{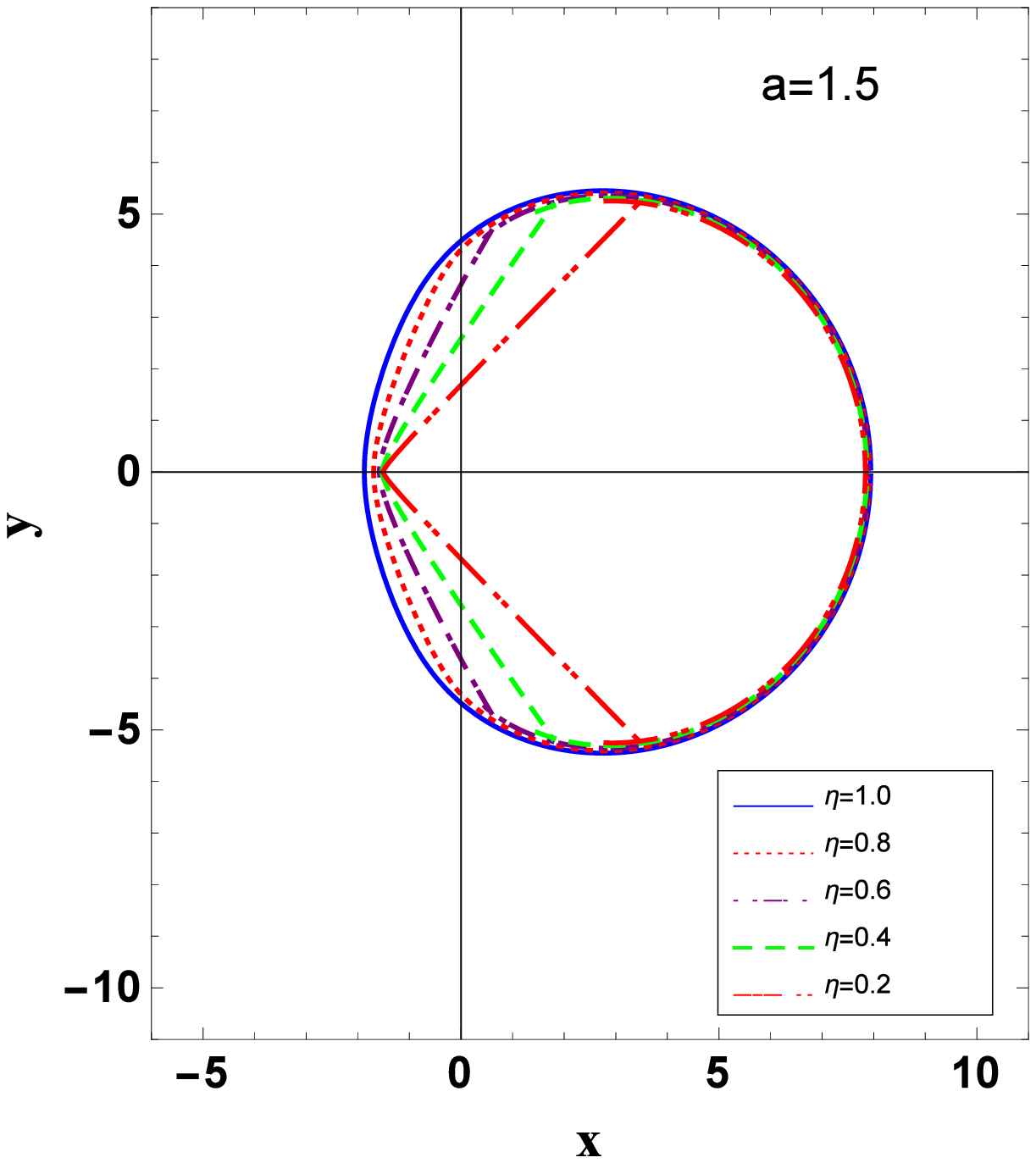}\includegraphics[width=7cm ]{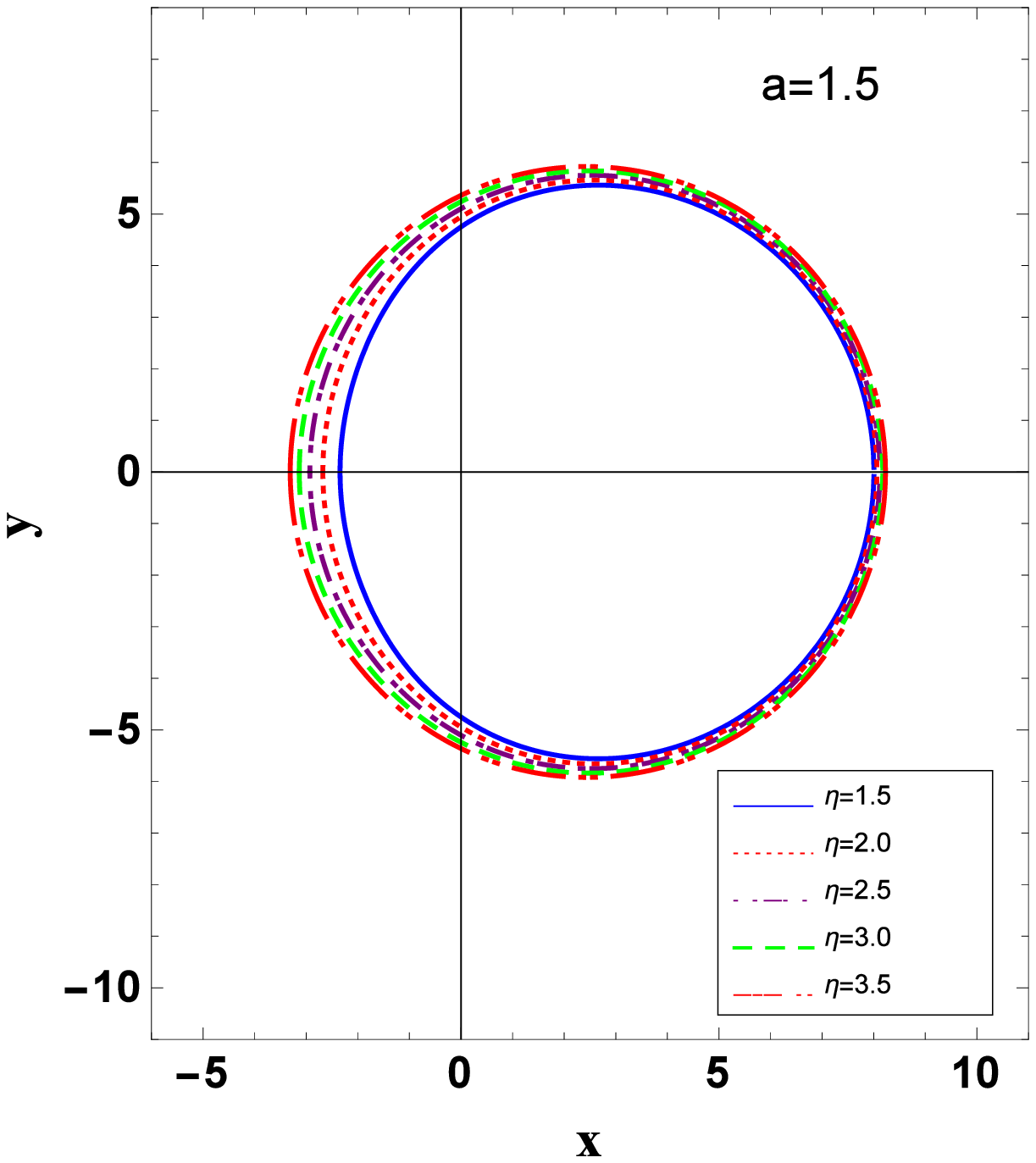}
\caption{The shadow of a Konoplya-Zhidenko rotating non-Kerr black hole with deformation parameter $\eta$ for the fixed $a=1.5$ lied in the range $a>\frac{2\sqrt{3}}{3}M$.  Here, an observer is situated at the origin of coordinates with the inclination angle $\theta_{0}=90\degree$.}
\label{a99}
\end{figure}

In the Konoplya-Zhidenko rotating non-Kerr black hole spacetime, one can obtain the position of photon's image in observer's sky
\begin{eqnarray}
\label{xd1}
x&=&-r_{0}\frac{p^{\hat{\phi}}}{p^{\hat{r}}}
=-r_{0}\frac{\Delta\sqrt{g_{rr}}L_{z}}{\sqrt{g_{\phi\phi}R(r_{0})}}, \nonumber\\
y&=&r_{0}\frac{p^{\hat{\theta}}}{p^{\hat{r}}}
=r_{0}\frac{\Delta\sqrt{g_{rr}}p_{\theta}}{\sqrt{g_{\theta\theta}R(r_{0})}},
\end{eqnarray}
if the observer is located at a distance $r=r_{0}$ and $\theta=\theta_{0}$. Considering that a real observer is far from the black hole, so we can take the limit $r_{0}\rightarrow\infty$, which yields
\begin{eqnarray}
\label{xd1w}
x=-\frac{\xi}{\sin \theta_{0}}, \;\;\;\;\;\;\;\;\;\;\;\;\;\;\;\;\;\;\;\;
y=\pm\sqrt{\sigma+a^{2}\cos^{2}\theta_{0}-\xi^{2}\cot^{2}\theta_{0}}.
\end{eqnarray}
\begin{figure}
\includegraphics[width=7cm ]{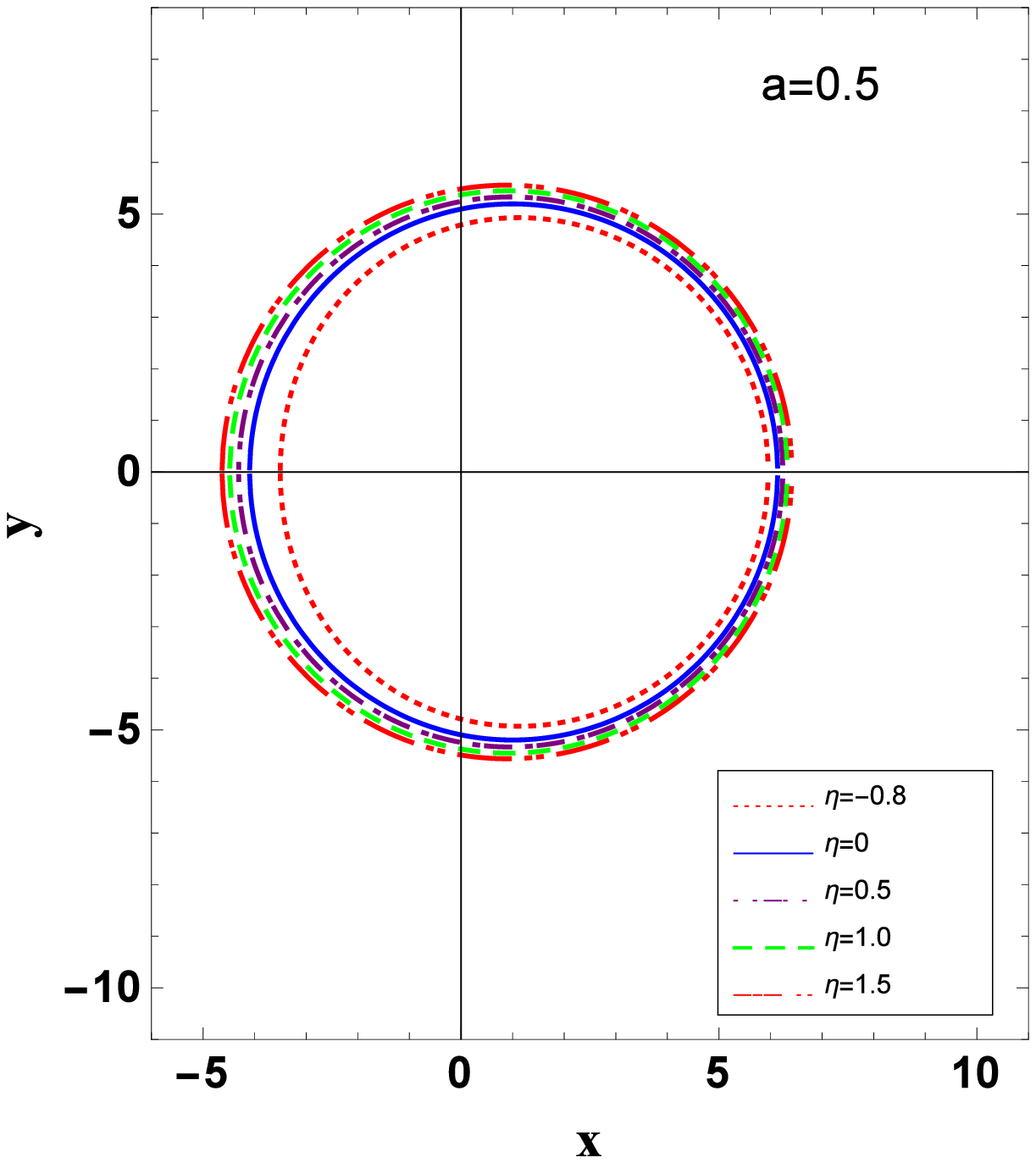}\includegraphics[width=7cm ]{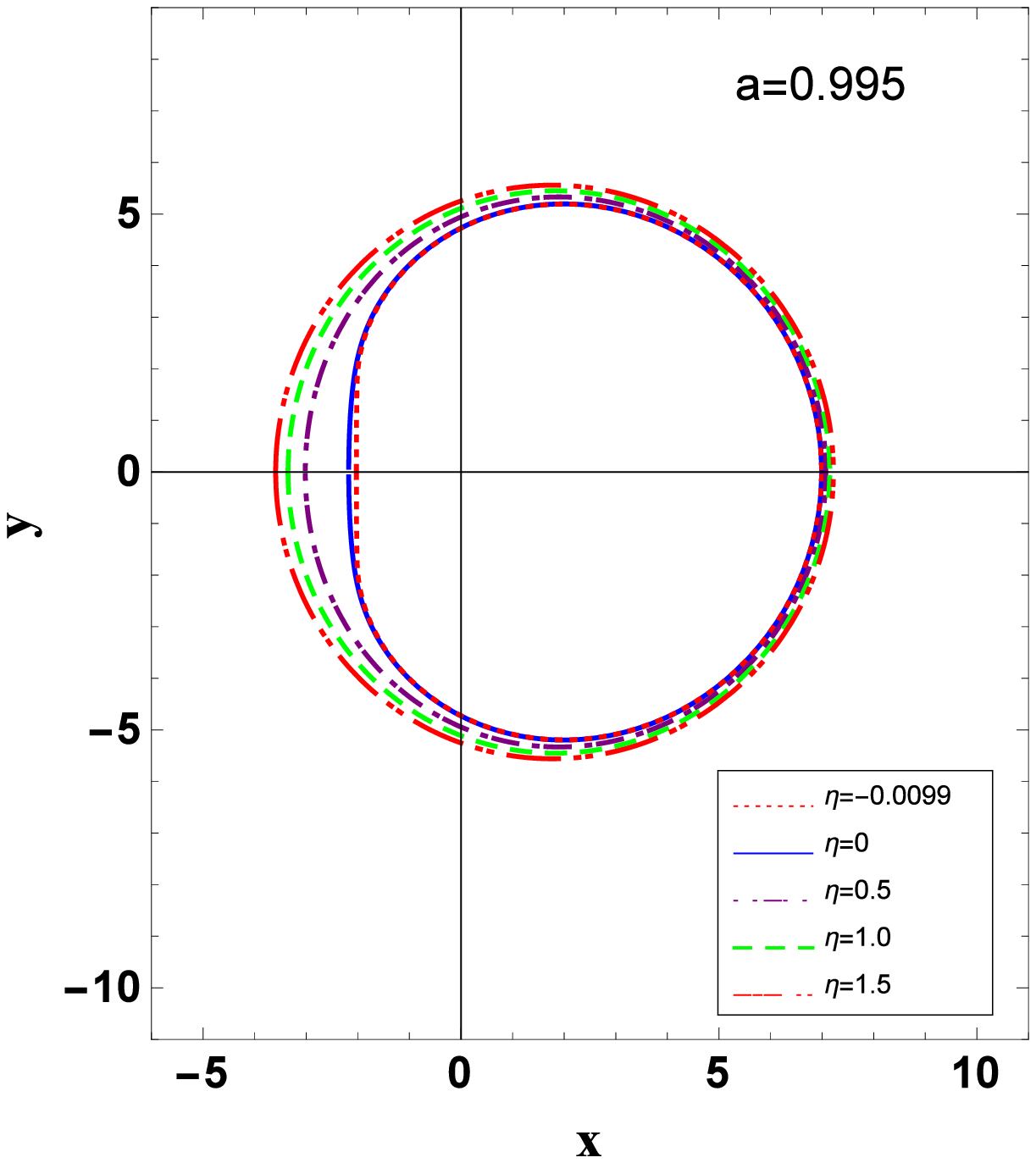}
\caption{The shadow of a Konoplya-Zhidenko rotating non-Kerr black hole with deformation parameter $\eta$ as the rotation parameter $a$ is the range $a<M$. The left and right panels correspond the cases with $a=0.5M$ and $a=0.995M$, respectively.  Here, an observer is situated at the origin of coordinates with the inclination angle $\theta_{0}=90\degree$. }
\label{a059}
\end{figure}
Together with Eqs.(\ref{pj}), (\ref{q}),  and (\ref{xd1w}), we
can study the properties of the shadow casted by a Konoplya-Zhidenko rotating non-Kerr black hole.
In Figs.(\ref{a059})-(\ref{a99}), we present the change of the black hole shadow with deformation parameter $\eta$ for the fixed $a$ as observer situates at the origin of coordinates with the inclination angle $\theta_{0}=90\degree$. The closed region enclosed by the curves is black hole shadow and the value of $\eta$ is chosen to ensure the existence of horizon so that the metric (\ref{xy}) describes geometry of a black hole. With the increase of the deformation parameter $\eta$, we find that the size of the shadow of  black hole increase for different $a$, while the shape of the shadow depends on the rotation parameter $a$. For the cases with $a<M$, for example, as $a=0.5M$ or $a=0.995M$, we find that the shape is a deformed circle for different $\eta$ and it is more deformed with the increase of the rotation parameter. In the case $a<M$, the D-shape shadow appears only if the values of $\eta$ approaches to the threshold value $\eta_{c1}$ and the rotation parameter $a$ is close to the mass of black hole $M$, which is similar to that of a usual Kerr black hole. However, with the increase of the deformation parameter $\eta$, the shape becomes less deformed for the Konoplya-Zhidenko rotating non-Kerr black hole. For the cases with $M<a<\frac{2\sqrt{3}}{3}M$, one can find that $\eta_{c1}>0$ and the condition for the existence of black hole horizon is the deformation parameter $\eta>0$. As the parameter $\eta$ is larger much than the threshold value $\eta_{c1}$,
we find from Fig.(\ref{a050}) that the black hole shadow still has a deformed circle silhouette in these cases. As the parameter $\eta$ decreases down to the value near $\eta_{c1}$, the shape of the black hole shadow changes gradually from a deformed circle silhouette to D-shape shadow, which is similar to those in the cases with $a<M$. However, the condition D-shape shadow appeared becomes only that the values of $\eta$ approaches to the threshold value $\eta_{c1}$ from positive direction. As $\eta$ decreases further and its value is below $\eta_{c1}$, we find that there is a distinct change for the black hole shadow and it becomes a cusp silhouette with small eye lashes. The non-smooth edge is the distinct feature of such kind of cusp shadows.
The cusp shadow of a rotating black hole is also found in the case of a high-hairy black hole where the null geodesics are not variable-separable \cite{fpos2}.
 With the increase of the deformation parameter $\eta$, the apex moves along left, but the angle of the cusp increases so that the shadow becomes less cuspidal.
\begin{figure}
 \subfigure[]{ \includegraphics[width=5.15cm ]{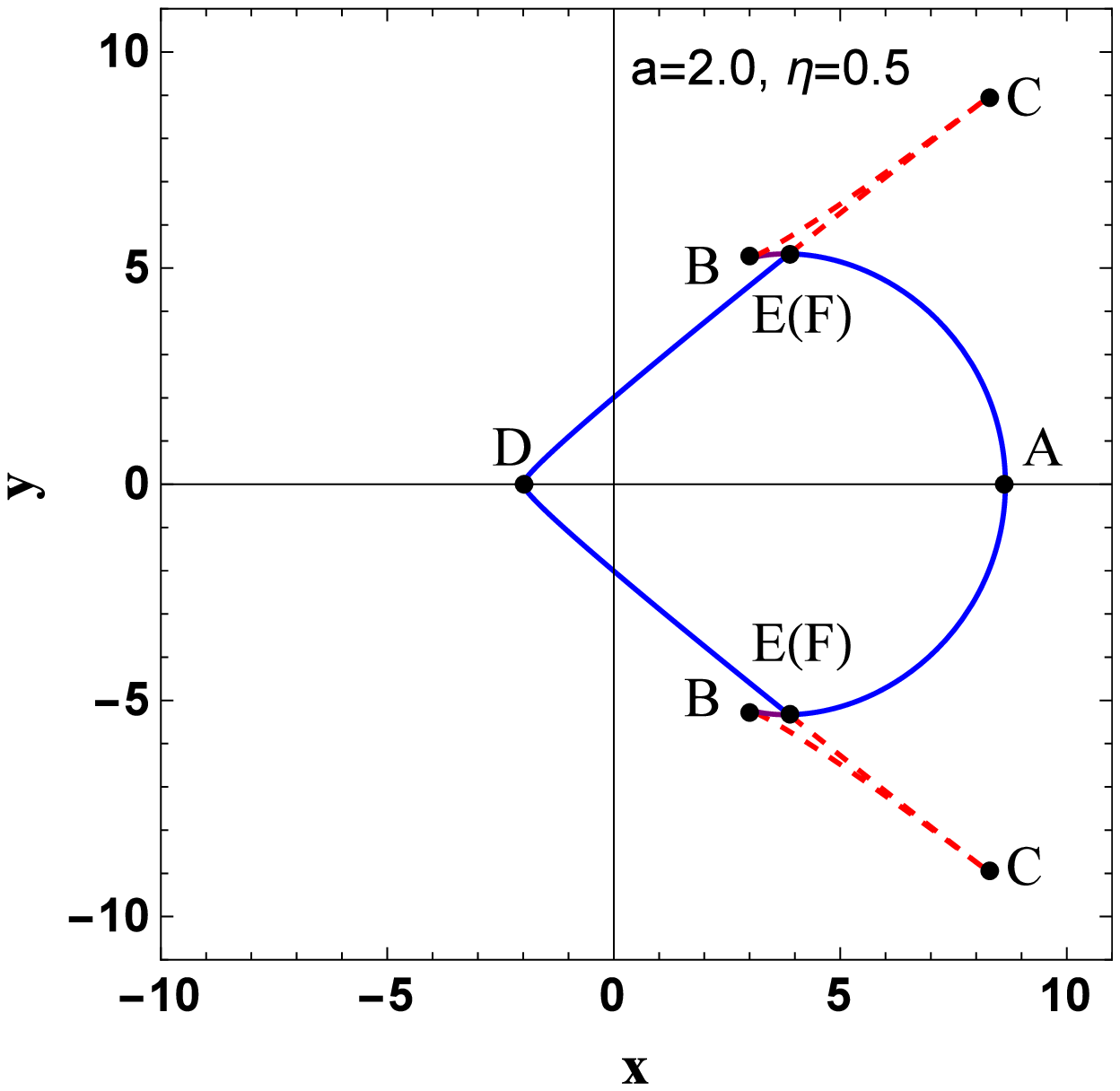} } \subfigure[]{ \includegraphics[width=5.5cm ]{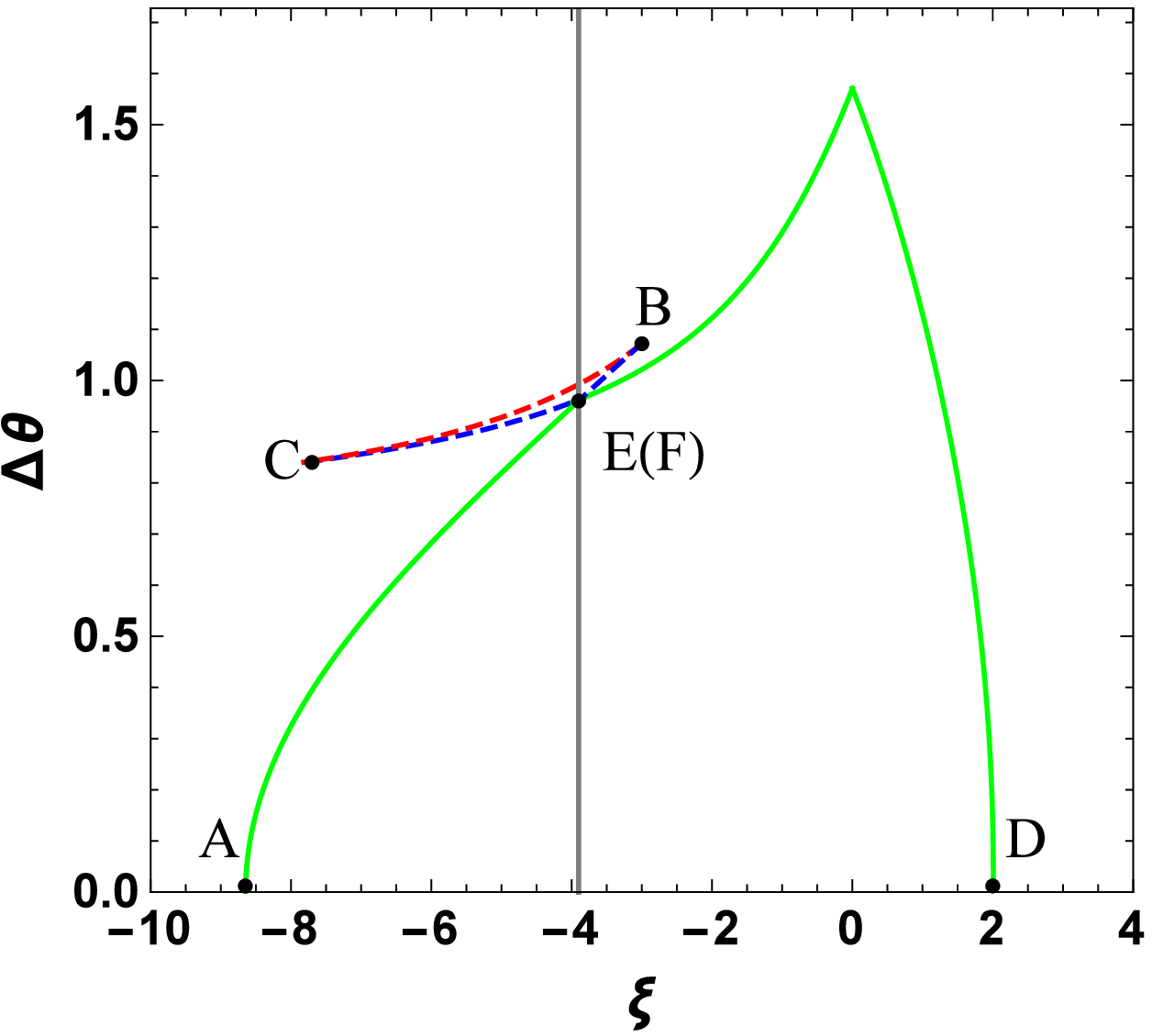}}\subfigure[]{ \includegraphics[width=5.3cm ]{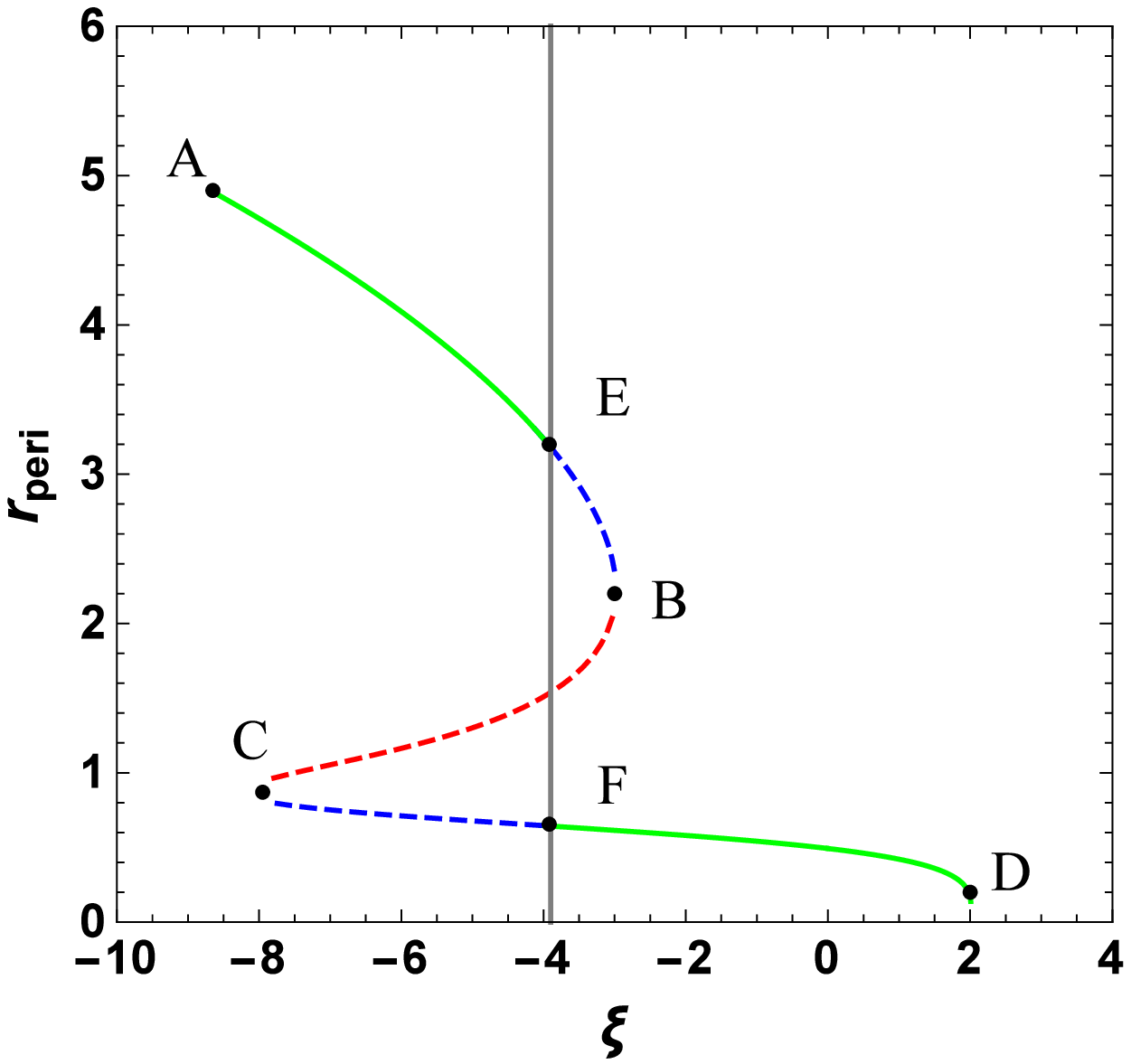}}
\caption{ The cusp shadow of a Konoplya-Zhidenko rotating non-Kerr black hole with $\theta_{0}=90\degree$ and FPOs for $a=2M, \eta=0.5$. In the left panel, the black hole shadow is described by the closed region enclosed by the solid curves. The middle and right panels correspond to the changes of $\Delta\theta$ and $r_{\text{peri}}$ for FPOs with $\xi$, respectively. }
\label{sab}
\end{figure}
As $a>\frac{2\sqrt{3}}{3}M$, the threshold value $\eta_{c1}$ is imaginary and the spacetime (\ref{xy}) possesses only a horizon for the positive $\eta$.
\begin{figure}
\subfigure[]{ \includegraphics[width=5.15cm ]{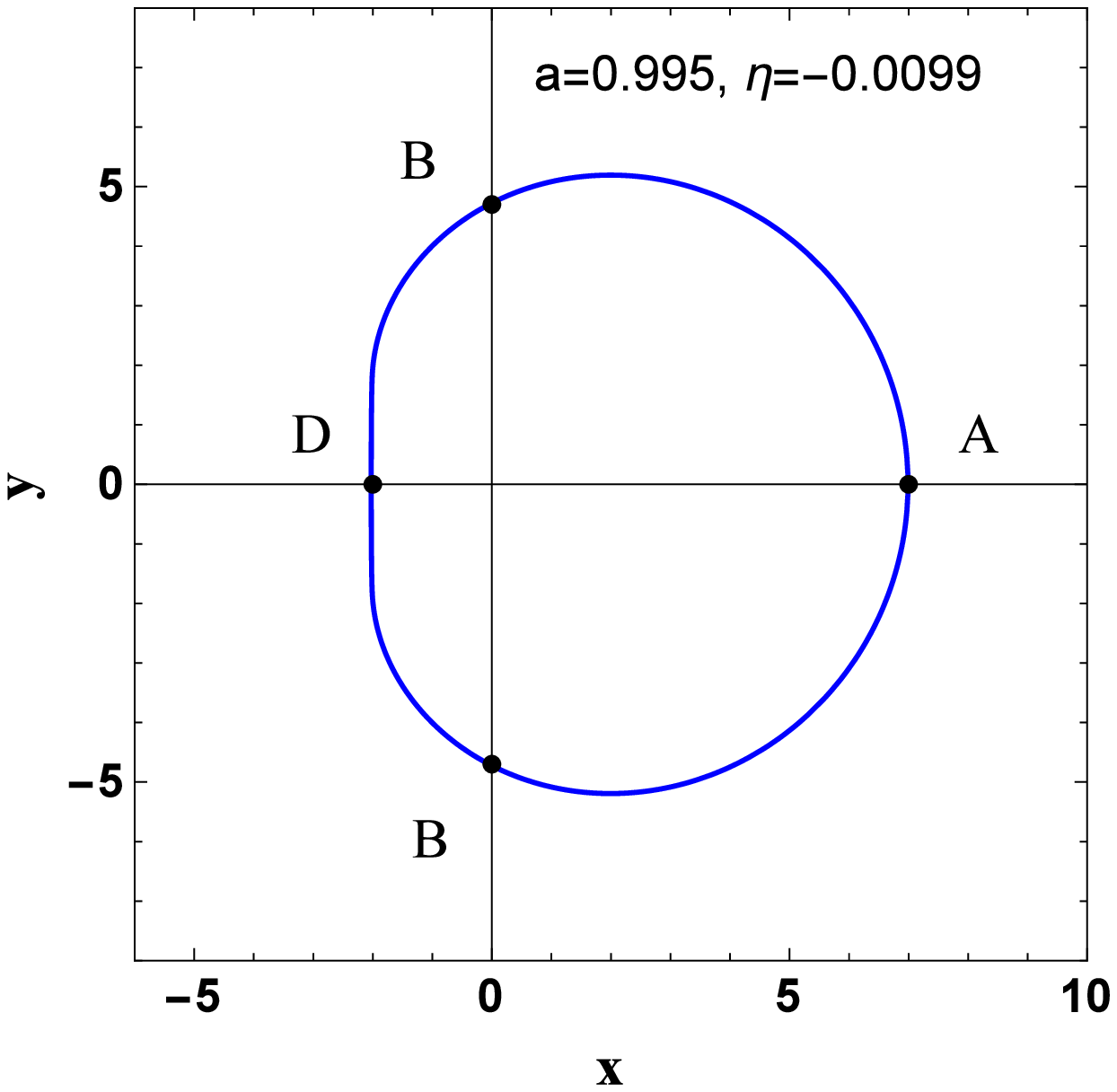}} \subfigure[]{ \includegraphics[width=5.5cm ]{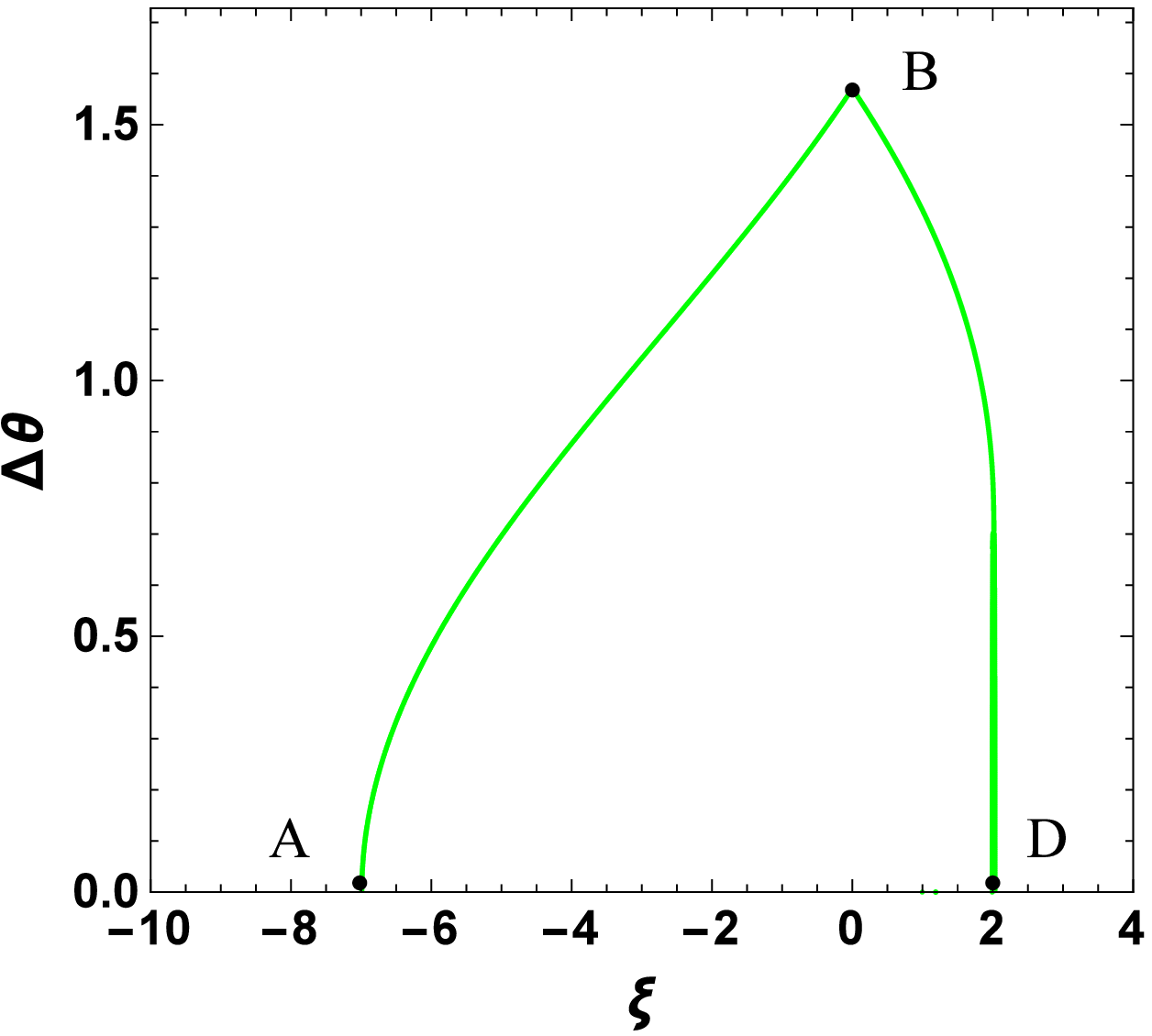}}\subfigure[]{ \includegraphics[width=5.3cm ]{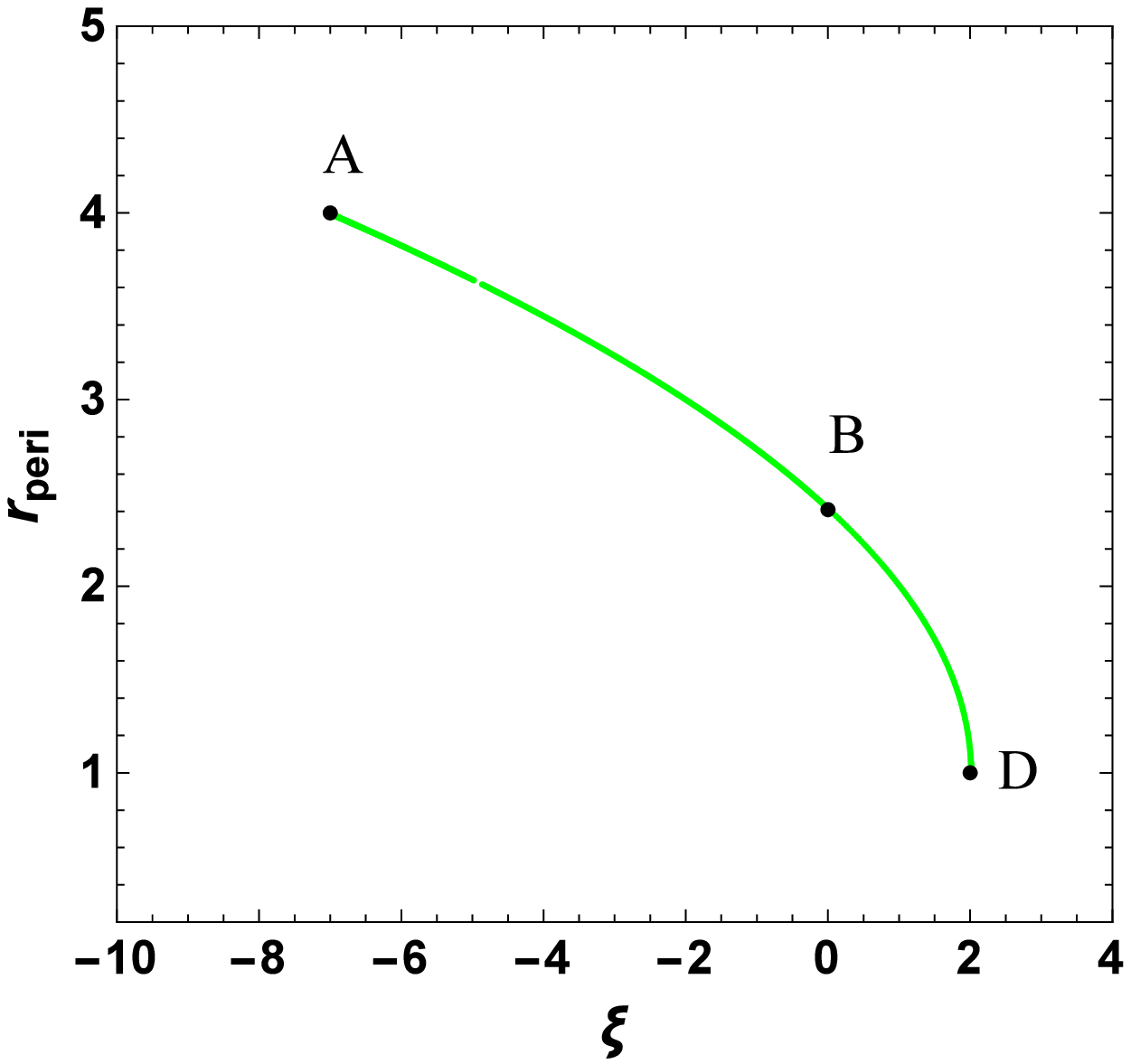}}
\caption{The D-shape shadow of a Konoplya-Zhidenko rotating non-Kerr black hole with $\theta_{0}=90\degree$ and FPOs for $a=0.995M, \eta=-0.0099$. In the left panel, the black hole shadow is described by the closed region enclosed by the solid curves. The middle and right panels correspond to the changes of $\Delta\theta$ and $r_{\text{peri}}$  for FPOs with $\xi$, respectively.}
\label{sab0}
\end{figure}
For the larger $\eta$, we find that the black hole shadow emerges as a deformed circle shape like in the previous cases, which is shown in Fig.(\ref{a99}). With the gradual decrease of the deformation parameter $\eta$, the shape of the shadow changes also from a deformed circle silhouette to a cusp shadow. However, the D-shape shadow does not appear in this case. In the previous discussion, we note that the D-shape shadow of a black hole emerges as the deformation parameter $\eta$ decreases down to the value near $\eta_{c1}$. The absence of D-shape shadow in the cases with $a>\frac{2\sqrt{3}}{3}M$ maybe be attributed to the nonexistence of a real threshold value $\eta_{c1}$. Moreover, the changes of the apex and  cusp angle with the deformation parameter $\eta$ are similar to those in the case of $M<a<\frac{2\sqrt{3}}{3}M$.

Finally, we discuss the formation of the cusp shadow casted by a Konoplya-Zhidenko rotating non-Kerr black hole as in Ref. \cite{fpos2}. Recently, it is pointed out by Pedro V. P. Cunha et al \cite{fpos2} that there exists non-planar bound photon orbits in some generic stationary, axisymmetric spacetimes, regardless of the integrability properties
of the photon motion. These non-planar bound photon orbits, were first proposed to explain the formation of a black hole shadow \cite{fpos2}. Actually, these orbits are generalization of circular photon orbits and they are also called as the fundamental photon orbits (FPOs) in the analysis on light bending by ultracompact objects \cite{fpos2}.
In Fig.(\ref{sab}),  we plot the change of quantities $\Delta\theta\equiv|\theta_{\text{max}}-\frac{\pi}{2}|$ and $r_{\text{peri}}$ with the impact parameter $\xi$ to display FPOs in the case with $a=2M$ and $\eta=0.5$ where the cusp shadow emerges for a Konoplya-Zhidenko rotating non-Kerr black hole. Here $\theta_{\text{max}}$  denote the maximal/minimal angular coordinate of a FPO, and $r_{\text{peri}}$ is the perimetral radius as a FPO crosses the equatorial plane.  From Fig.(\ref{sab}), one can find that the circular photon orbits (i.e., light rings) are connected by a continuum of
FPOs marked with the capital letters A-F, which can be split into one stable branch and two unstable ones as in the case of a rotating hairy black hole \cite{fpos2}.  It is shown that only a part of unstable FPOs ( the green line parts shown in Fig.(\ref{sab})) determine the boundary of shadow. Obviously, one can find  there exist a swallow-tail shape pattern related to FPOs in the $\xi-\Delta\theta$ plane, which yields a jump occurred at the FPOs E and F.
The discontinuity in the size of these orbits (i.e., $r_{\text{peri}(E)}>r_{\text{peri}(F)}$) originating from this jump
induces further the emergence of the cusp in the shadow. In order to make a comparison, we also plot FPOs ( marked with the capital letters A, B, and D ) in Fig.(\ref{sab0}) for the case with $a=0.995$ and $\eta=-0.0099$ in which the black hole shadow is a D-shape silhouette  rather than a cusp one. It is easy to find that the swallow-tail shape pattern related to FPOs does not appear in the $\xi-\Delta\theta$ plane and the corresponding the perimetral radius of FPOs crossing  the equatorial plane is continuous in this case, which leads to the disappearance of cusp shadow of black hole. Our result show that although
the dynamical system about photon motion is integrable in a Konoplya-Zhidenko rotating non-Kerr black hole spacetime, the rich spacetime properties arising from the deformation parameter $\eta$ yields a series of significant patterns for black hole shadow.

\section{Summary}

In this paper we have studied the shadow casted by a Konoplya-Zhidenko rotating non-Kerr black hole with an extra deformation parameter. We find that the deformed parameter together with the rotation parameter affects spacetime structure and the black hole shadow. With the increase of the deformation parameter, the size of the shadow of  black hole increase and its shape becomes more rounded for arbitrary rotation parameter. The condition the D-shape shadow of black hole emerged depends on the value of $a$.
In the case $a<M$, the D-shape shadow appears only if the values of $\eta$ approaches to the threshold value $\eta_{c1}$ and the rotation parameter $a$ is close to the mass of black hole $M$. For the case $M<a<\frac{2\sqrt{3}}{3}M$, the condition D-shape shadow appeared becomes only that the values of $\eta$ approaches to the threshold value $\eta_{c1}$ from positive direction.  For the case $a>\frac{2\sqrt{3}}{3}M$, there is no D-shape shadow for the black hole.
Moreover,  we find that the black hole shadow has a cusp shape with small eye lashes in the cases with $a>M$.
With the increase of the deformation parameter $\eta$, the cusp apex moves along left, but the angle of the cusp increases, which means that the shadow becomes less cuspidal in this case. Finally, we analyse the FPOs and
discuss further the formation of the cusp shadow casted by a Konoplya-Zhidenko rotating non-Kerr black hole. Our result show that the richer spacetime properties arising from the deformation parameter $\eta$ yields a series of significant patterns for the shadow casted by a Konoplya-Zhidenko rotating non-Kerr black hole.

\section{\bf Acknowledgments}

This work was partially supported by the Scientific Research
Fund of Hunan Provincial Education Department Grant
No. 17A124. J. Jing's work was partially supported by
the National Natural Science Foundation of China under
Grant No. 11475061.


\begin{thebibliography}{99}
\baselineskip=0.6cm

\bibitem{sha1} J. L. Synge, \textit{The Escape of Photons from Gravitationally Intense Stars}, Mon. Not. Roy. Astron. Soc. {\bf131}, 463 (1966).
\bibitem{sha2} J. M. Bardeen, in \textit{Black Holes (Les Astres Occlus)}, edited by C. DeWitt and B. DeWitt (Gordon and Breach, New York, 1973), p. 215-239.


\bibitem{sha3}S. Chandrasekhar, \textit{The Mathematical Theory of Black Holes} (Oxford University Press, New York, 1992).

\bibitem{sha4} H. Falcke, F. Melia, and E. Agol, \textit{Viewing the Shadow of the Black Hole at the Galactic Center}, Astrophys. J. {\bf528}, L13 (2000), arXiv:astro-ph/9912263.
\bibitem{sha5} A. de Vries,  \textit{The apparent shape of a rotating charged black hole, closed photon orbits and the bifurcation set $A_4$}, Class. Quant. Grav. {\bf17},123 (2000).
\bibitem{sha6} R. Takahashi, \textit{Shapes and Positions of Black Hole Shadows in Accretion Disks and Spin Parameters of Black Holes}, Astrophys. J. {\bf611}, 996 (2004), arXiv:astro-ph/0405099.
\bibitem{sha7} C. Bambi and K. Freese, \textit{Apparent shape of super-spinning black holes}, Phys. Rev. D {\bf 79}, 043002 (2009), arXiv:0812.1328 [astro-ph];

Z. Li and C. Bambi, \textit{Measuring the Kerr spin parameter of regular black holes from their shadow}, J. Cosmol. Astropart. Phys. {\bf1401}, 041(2014),arXiv:1309.1606[gr-qc].





\bibitem{sha9} K. Hioki and K. I. Maeda, \textit{Measurement of the Kerr Spin Parameter by Observation of a Compact Object's Shadow}, Phys. Rev. D {\bf80}, 024042 (2009), [arXiv:0904.3575 [astro-ph.HE]].

\bibitem{sha10} L. Amarilla, E. F. Eiroa, and G. Giribet, \textit{Null geodesics and shadow of a rotating black hole in extended Chern-Simons modified gravity}, Phys. Rev. D {\bf81}, 124045 (2010), [arXiv:1005.0607[gr-qc]].

\bibitem{sha11}L. Amarilla and E. F. Eiroa, \textit{Shadow of a rotating braneworld black hole}, Phys. Rev. D {\bf85}, 064019 (2012), [arXiv:1112.6349 [gr-qc]].
\bibitem{sha12} A. Yumoto, D. Nitta, T. Chiba, and N. Sugiyama,\textit{Shadows of Multi-Black Holes: Analytic Exploration}, Phys. Rev. D {\bf86}, 103001 (2012), [arXiv:1208.0635 [gr-qc]].
\bibitem{sha13} L. Amarilla and E. F. Eiroa,  \textit{Shadow of a Kaluza-Klein rotating dilaton black hole}, Phys. Rev. D {\bf87}, 044057 (2013), [arXiv:1301.0532 [gr-qc]].
\bibitem{sha14} P. G. Nedkova, V. Tinchev, and S. S. Yazadjiev, \textit{The Shadow of a Rotating Traversable Wormhole}, Phys. Rev. D {\bf88}, 124019 [arXiv:1307.7647[gr-qc]];

    V. K. Tinchev and S. S. Yazadjiev, \textit{Possible imprints of cosmic strings in the shadows of galactic black holes}, Int. J. Mod. Phys. D {\bf23},  1450060 (2014),  [arXiv:1311.1353[gr-qc]];
\bibitem{sha15} S. W. Wei and Y. X. Liu,\textit{Observing the shadow of Einstein-Maxwell-Dilaton-Axion black hole}, J. Cosmol. Astropart. Phys. {\bf11}, 063 (2013).
\bibitem{sha16} V. Perlick, O. Y. Tsupko, and G. S. Bisnovatyi-Kogan, \textit{Influence of a plasma on the shadow of a spherically symmetric black hole}, Phys. Rev. D {\bf92}, 104031 (2015).
\bibitem {sb1} Y. Huang, S. Chen, and J. Jing, \textit{Double shadow of a regular phantom black hole as photons couple to Weyl tensor}, Eur. Phys. J. C {\bf76}, 594 (2016).


\bibitem{sha17} Z. Younsi, A. Zhidenko, L. Rezzolla, R. Konoplya and Y. Mizuno, \textit{A new method for shadow calculations: application to parameterised axisymmetric black holes}, Phys. Rev. D {\bf94}, 084025 (2016), arXiv:1607.05767.

\bibitem{sha18} A. Bohn, W. Throwe, F. Hbert, K. Henriksson, and D. Bunandar, \textit{What does a binary black hole merger look like?}, Class. Quantum Grav. {\bf32}, 065002 (2015), arXiv: 1410.7775.

\bibitem{sha19} A. Abdujabbarov, M. Amir, B. Ahmedov, S. Ghosh, \textit{Shadow of rotating regular black holes},  Phys. Rev. D{\bf93}, 104004 (2016);

    A. Abdujabbarov, L. Rezzolla, B. Ahmedov, \textit{A coordinate-independent characterization of a black hole shadow}, Mon. Not. R. Astron. Soc. {\bf 454}, 2423 (2015);

     F. Atamurotov, B. Ahmedov, A. Abdujabbarov, \textit{Optical properties of black hole in the presence of plasma: shadow},
     Phys. Rev D {\bf92}, 084005 (2015);

     A. Abdujabbarov, F. Atamurotov, N. Dadhich, B. Ahmedov, Z. Stuchl\'{i}k, \textit{Energetics and optical properties of 6-dimensional rotating black hole in pure Gauss-Bonnet gravity}, Eur. Phys. J. C {\bf75}, 399 (2015);

     F. Atamurotov, A. Abdujabbarov, B. Ahmedov, \textit{Shadow of rotating non-Kerr black hole}, Phys. Rev D {\bf88}, 064004 (2013).




\bibitem{sw} P. V. P. Cunha, C. Herdeiro, E. Radu and H. F. Runarsson, \textit{Shadows of Kerr black holes with scalar hair}, Phys. Rev. Lett. {\bf115}, 211102 (2015), arXiv:1509.00021;
\bibitem{swo} P. V. P. Cunha, C. Herdeiro, E. Radu and H. F. Runarsson, \textit{Shadows of Kerr black holes with and without scalar hair}, Int. J. Mod. Phys. D {\bf25}, 1641021 (2016), arXiv:1605.08293.

\bibitem{astro}F. H. Vincent, E. Gourgoulhon, C. Herdeiro and E. Radu, \textit{Astrophysical imaging of Kerr black holes with scalar hair}, Phys. Rev. D {\bf94}, 084045 (2016), arXiv:1606.04246.


\bibitem{chaotic} P. V. P. Cunha, J. Grover, C. Herdeiro, E. Radu, H. Runarsson, and A. Wittig, \textit{Chaotic lensing around boson stars and Kerr black holes with scalar hair}, Phys. Rev. D {\bf94}, 104023 (2016).

\bibitem{binary} J. O. Shipley, and S. R. Dolan, \textit{Binary black hole shadows, chaotic scattering and the Cantor set}, Class. Quantum Grav. {\bf33}, 175001 (2016).


\bibitem{harip1}C. Herdeiro, E. Radu, and H. Runarsson, \textit{Kerr black holes with Proca hair}, Class. Quantum Grav. {\bf33}, 154001 (2016), arXiv: 1603.02687.

\bibitem{fpos2}P. V. P. Cunha, C. Herdeiro, and E. Radu, \textit{Fundamental photon orbits: black hole shadows and spacetime instabilities}, Phys. Rev. D {\bf96}, 024039 (2017).

\bibitem{t3} C. M. Will,\textit{The Confrontation between General Relativity and Experiment},  Living Rev. Rel. {\bf17} (2014) 4.



\bibitem{kz} R. Konoplya, A. Zhidenko,\textit{Detection of gravitational waves from black holes: Is there a window for alternative theories?}, Phys. Lett. B {\bf756},350 (2016).

\bibitem{RLs}R. Konoplya, L. Rezzolla and A. Zhidenko, \textit{General parametrization of axisymmetric black holes in metric theories of gravity}, Phys. Rev. D {\bf93}, 064015 (2016).
\bibitem{sy} S. Wang, S. Chen, J. Jing, \textit{Strong gravitational lensing by a Konoplya-Zhidenko rotating non-Kerr compact object}, J. Cosmol. Astropart. Phys. {\bf11}, 020 (2016).


\bibitem{GKt02} Y. Ni, J. Jiang and C. Bambi, \textit{Testing the Kerr metric with the iron line and the KRZ parametrization}, J. Cosmol. Astropart. Phys. {\bf09},  014 (2016).
\bibitem{GKt01} C. Bambi and S. Nampalliwar,\textit{Quasi-periodic oscillations as a tool for testing the Kerr metric: A comparison with gravitational waves and iron line},  Europhys. Lett. {\bf116}, 30006 2016.

\bibitem{hamin} B. Carter, \textit{Global structure of the Kerr family of gravitational fields}, Phys. Rev. {\bf174}, 1559 (1968).

\bibitem{hamin1} M. Walker and R. Penrose, \textit{On quadratic first integrals of the geodesic equations for type $\{ 22 \}$ spacetimes}, Commun. math. Physics {\bf 18}, 265 (1970).

\bibitem{zero1}V. P. Frolov and I. D. Novikov, \textit{Black Hole Physics: Basic concepts and new developments}, Kluwer Academic Publishers, 1998.
\bibitem{swo7} T. Johannsen, \textit{Photon Rings around Kerr and Kerr-like Black Holes}, Astrophys. J. {\bf777}, 170, (2013).










\end{thebibliography}
\end{document}